\documentclass[iop]{emulateapj}
\usepackage{apjfonts}
\shorttitle{Lithium in V1369 Cen}
\shortauthors{Izzo et al.}

\begin{document}


\title{Early optical spectra of nova V1369 Cen show presence of Lithium}


\author{Luca Izzo}
\affil{Physics Department, Sapienza University of Rome, I-00185; ICRANET, Piazza della Repubblica 10, 65122, Pescara, Italy} 
	\email{luca.izzo@icra.it - luca.izzo@gmail.com}

\author{Massimo Della Valle}
\affil{INAF, Osservatorio Astronomico di Capodimonte, salita Moiariello 16, 80131, Napoli; ICRANET, Piazza della Repubblica 10, 65122, Pescara, Italy}

\author{Elena Mason}
\affil{INAF, Osservatorio Astronomico di Trieste, via G.B. Tiepolo 11, I-34143 Trieste, Italy}

\author{Francesca Matteucci}
\affil{Dipartimento di Fisica, Sezione di Astronomia, Universit\'a di Trieste, via G.B. Tiepolo 11, I-34143 Trieste, Italy}

\author{Donatella Romano}
\affil{INAF, Osservatorio Astronomico di Bologna, via Ranzani 1, 40127, Bologna, Italy}

\author{Luca Pasquini}
\affil{ESO, Karl-Schwarzschild-Strasse 2, 85748 Garching bei Munchen, Germany}

\author{Leonardo Vanzi}
\affil{Department of Electrical Engineering and Center of Astro Engineering, PUC-Chile, 
	Avenida Vicu$\tilde{n}$a Mackenna 4860, Santiago, Chile}

\author{Andres Jordan}
\affil{Institute of Astrophysics and Center of Astro Engineering, PUC-Chile, 
	Avenida Vicu$\tilde{n}$a Mackenna 4860, Santiago, Chile}

\author{Jos\'e Miguel Fernandez, Paz Bluhm, Rafael Brahm, Nestor Espinoza}
\affil{Institute of Astrophysics, PUC-Chile, Avenida Vicu$\tilde{n}$a Mackenna 4860, Santiago, Chile}

\and 

\author{Robert Williams}
\affil{Space Telescope Science Institute, 3700 San Martin Drive, Baltimore, MD 21218, USA}



\begin{abstract}
We present early high resolution spectroscopic observations of the nova V1369 Cen.  We have detected an absorption feature at 6695.6 \AA\, that we have identified as blue--shifted  $^7$Li I $\lambda$6708 \AA. The absorption line, moving at -550 km/s, was observed in five high-resolution spectra of the nova obtained at different epochs. On the basis of the intensity of this absorption line we infer that a single nova outburst can inject in the Galaxy $M_{Li} =$ 0.3 - 4.8 $\times 10^{-10}$ M$_{\odot}$. Given the current estimates of Galactic nova rate, this amount is sufficient to explain the puzzling origin of the overabundance of Lithium observed in young star populations.
\end{abstract}





\section{Introduction}

The light elements Deuterium, 3-Helium, 4-Helium, and 7-Lithium are synthesized in non-negligible amounts during the first few minutes of the initial cosmic expansion \citep{KolbTurner}. For Big-Bang Nucleosynthesis in the standard model of cosmology and particle physics (SBBN), the predicted abundances for these elements depends only on one parameter, the baryon-to-photon density ratio. Recently, Planck data have determined the baryon density to excellent precision, leading to a primordial lithium abundance in the range $A(Li) = 2.66-2.73$\footnote{$A(Li) = log_{10} (N_{^7Li}/N_H) + 12$} \citep{Coc2014}. This value is significantly larger than $A(Li) \sim 2.1-2.3$ obtained for old metal-poor ([Fe/H] $ \leq - 1.4$) halo stars, whose distribution in the Lithium abundance -- metallicity diagram is almost flat, and defines the so--called ''Spite plateau'' \citep{Spite1982,Bonifacio2007}. These stars were long thought to share a common, primordial, Li abundance. The discrepancy may be explained by diffusion and turbulent mixing in the stellar interiors \citep{Korn2006}, which corrodes the primordial Li abundance, and/or by a non-standard BBN scenario \citep{Iocco2009}. Yet, another puzzle exists. The abundance of Lithium observed in the upper envelope of young metal-rich ([Fe/H]\footnote{[Fe/H] = $log_{10} \left(\frac{N_{Fe}}{N_H}\right)_{Nova} - log_{10} \left(\frac{N_{Fe}}{N_H}\right)_{Sun}$} $ > - 1.4$) stars traces a growth of the Li abundance from the Spite plateau value to the meteoritic value, $A(Li) = 3.26 \pm 0.05$ \citep{Lodders2009} and even higher, which clearly indicates that lithium enrichment mechanisms must occur on Galactic scales. 

In recent decades several potential Li producers have been proposed on theoretical grounds. Among the possible lithium farms in the Galaxy, the most plausible astrophysical sources are represented by asymptotic giant branch (AGB) stars \citep{Iben1973}, novae \citep{Starrfield1978} and Galactic cosmic ray spallation (via fragmentation of material due to impact of accelerated protons, leading to expulsion of nucleons, \citealp{Lemoine1998}). Galactic chemical evolution models \citep{DAntonaMatteucci1991,Romano2001,Travaglio2001,Prantzos2012}, show a convincing match with observations, when all lithium producers described above are included.

The main channel for thermonuclear production of lithium inside stars is based on a particular sequence of events, known as hot bottom burning and beryllium transport mechanism \citep{CameronFowler1971}.  At first there is formation of $^7$Be via the reaction $^3$He $+ \alpha$, which requires large temperatures $T \geq 10^7 K$, and occurs in the deep interior of stars. Outward convection mechanisms concurrently must transport the newly formed $^7$Be to external cooler regions, where $T \approx 10^6 K$. Here, beryllium is able to capture electrons (free and bound), giving rise to $^7$Li  with the additional formation of 0.86 MeV neutrinos. More recent computations \citep{Ventura2010} suggest that a large abundance of $^7$Li can thus be achieved in AGB stars, particularly with mass larger than $7 M_{\odot}$. However, large lithium yields and, consequently, a significant enrichment in Li of the surrounding interstellar medium, can be obtained from these sources only if extensive mass loss is associated with the phases of maximum Li production \citep{Romano2001}. 

In nova systems the abundance of lithium is related to the thermonuclear runaway (TNR) scenario, which accounts for the physical explanation of  $^7$Li and its ejection from the binary system. In TNRs the expanding gas reaches velocities of the order of 400-4000 km/s, which results in escape from the system, and enrichment of the interstellar medium with elements synthesized in the nova outburst \citep{JoseHernanz1998}. This scenario remains uncertain because of  the lack of direct detection of $^7$Li during nova outburst, although its presence was proposed long ago \citep{Starrfield1978}. Indeed, $^7$Li is easily destroyed by proton fusion at temperatures greater than $2.6 \times 10^6 K$, giving rise to two helium atoms. For this reason, $^7$Li detection in stars must focus on the external regions, where the temperature is sufficiently cool to allow non-depletion of $^7$Li. 

In the literature there have been reported only upper limits to the abundance of $^7$Li in selected novae, e.g., HR Del, IV Cep, and NQ Vul \citep{Friedjung1979}, while direct detections of lithium in symbiotic novae (e.g. T CrB, RS Oph and V407 Cyg) are characterized by velocities of a few dozens km/s, which implies that lithium is not associated with the WD ejecta, but rather originates in the secondary giant star \citep{Wallerstein2008,Brandi2009,Shore2011b}. 

Recently \citet{Tajitsu2015} have announced detection of $^7$Be features in the near-UV late spectra of V339 Del (Nova Delphini 2013). Indeed, the Cameron-Fowler mechanism is expected to occur during the TNR \citep{Starrfield1978}, leading to an over--production of beryllium at the epoch of the nova outburst, with a consequent decay to lithium after $\sim$ 50 days. The amount of Be observed is larger than typical values deduced from theoretical predictions for CO novae from thermonuclear runaway models \citep{Truran1981} representing the very early phases of the nova outburst. 

\begin{figure*}
\center
\includegraphics[width=0.49\textwidth]{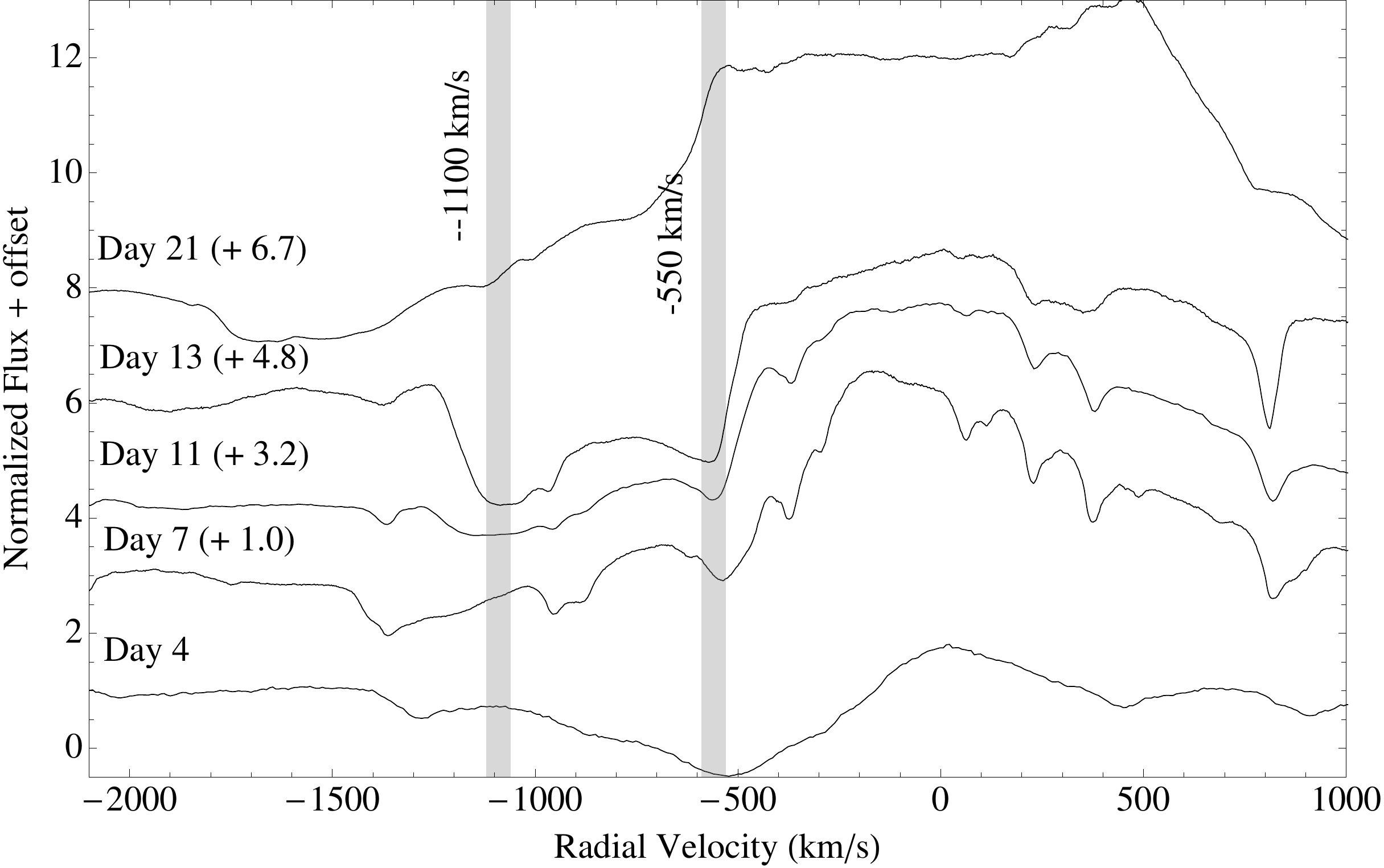}
\includegraphics[width=0.485\textwidth]{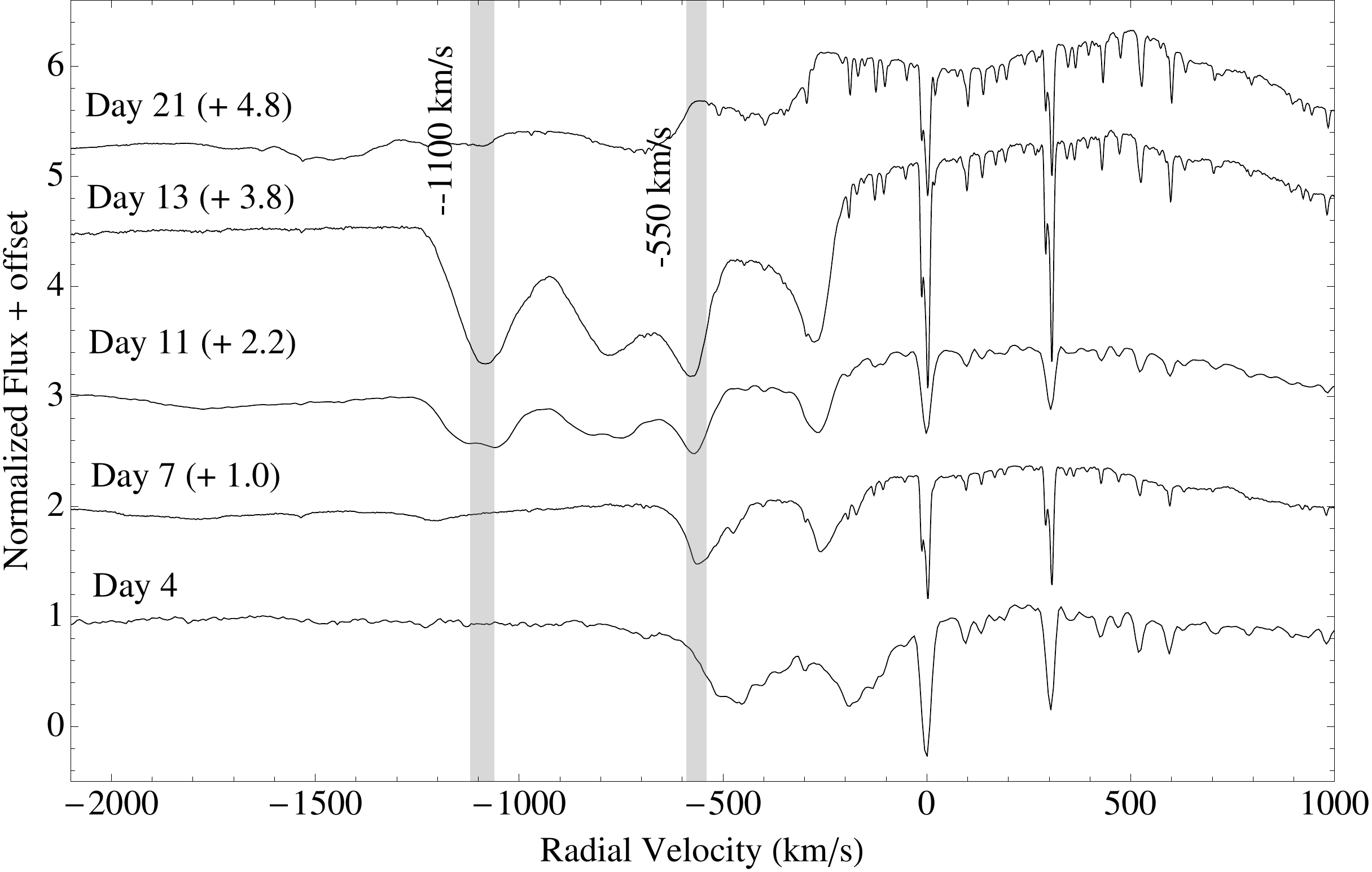}\\
\caption{The evolution of the radial velocity systems observed through the P-Cygni profiles in H$\beta$ (left panel - centered at $\lambda_{H\beta} = 4861.32$) and Na I D2 (right panel - $\lambda_{NaI} = 5889.93$) lines in the first three weeks of the V1369 Cen outburst. Expansion velocities of -550 and -1100 ($\pm 30$ km/s) are marked with gray rectangles. Note the appearance of multiple expanding systems with increasing time. In particular, the absence in the spectra of day 4 and day 7 of the component at higher velocities ($v_{exp} = -1350 $ km/s) in the Sodium P-Cygni profile, which is however observed in H$\beta$.
}
\label{fig:no1}
\end{figure*}

\section{Observations of $^7$Lithium I 6708 \AA}

We report here the detection of $^7$Li I 6708 \AA\, in the early (first three weeks) high-resolution spectra of the ``slow'' ($t_2 = 40 \pm 1$ days,  where $t_2$ corresponds to the time during which the nova declines by 2 magnitudes) nova V1369 Cen (Nova Cen 2013), when the nova was in the optically thick phase\footnote{http://lucagrb.altervista.org/research/lightcurve\_NCen\_1.pdf - courtesy AAVSO}.  Optical high-resolution spectral observations started 4 days after the initial outburst.  Early spectral data were obtained on days 7, 13, and 21 with the Fiber-fed Extended Range Optical Spectrograph (FEROS - R $\sim$ 48000) mounted on the MPG 2.2m telescope located on La Silla \citep{Kaufer1999}, and on days 4, 11, 16, and 18 with the PUC High Echelle Resolution Optical Spectrograph (PUCHEROS - R $\sim$ 20000) mounted on the ESO 0.5m telescope located at the Pontificia Universidad Catolica Observatory in Santiago \citep{Vanzi2012}. The spectra are all characterized by the presence of bright Balmer and typical Fe II emission lines, which suggests that the nova was engulfed in its ''iron curtain'' phase \citep{ShoreBASI}. The presence of many absorption features in the range 3700-4600 \AA\, in the first two weeks (days 4, 7, 11, and 13) complicates the identification of the most common transitions detected in the optically thick phase of novae, which are identified from their respective P--Cygni absorptions. We identify multiple expanding velocities for each transition : in the first week (days 4 and 7) we measure from the Balmer lines, O I 7773-7, 8446 \AA, and Fe II (multiplet 42) lines two expanding systems with mean velocities of $\sim$ -550 and -1350 (with a maximum value of -1400) km/s, while for Na I we see only the system expanding at $v_{exp} \sim -550$ km/s. In the second week (days 11 and 13) we identify two prominent expanding components at $v_{exp} \sim$ -550 and -1100 km/s from P-Cygni profiles of all main transitions (Balmer, O I, Fe II and Na I D), while an additional faster component is observed at $v_{exp} \sim -1900$ km/s in H$\beta$ and H$\alpha$ lines alone. In subsequent spectra  (day $\geq$21), we identify a broad and structured absorption at $v_{exp} \approx -1600$ km/s and two additional components at $\sim$ -550 and -750 km/s. The evolution of the observed P-Cygni absorptions of H$\beta$  and Na I D2 is shown in Fig. \ref{fig:no1}.
 
\begin{figure}[!t]
\center
\includegraphics[width=0.49\textwidth]{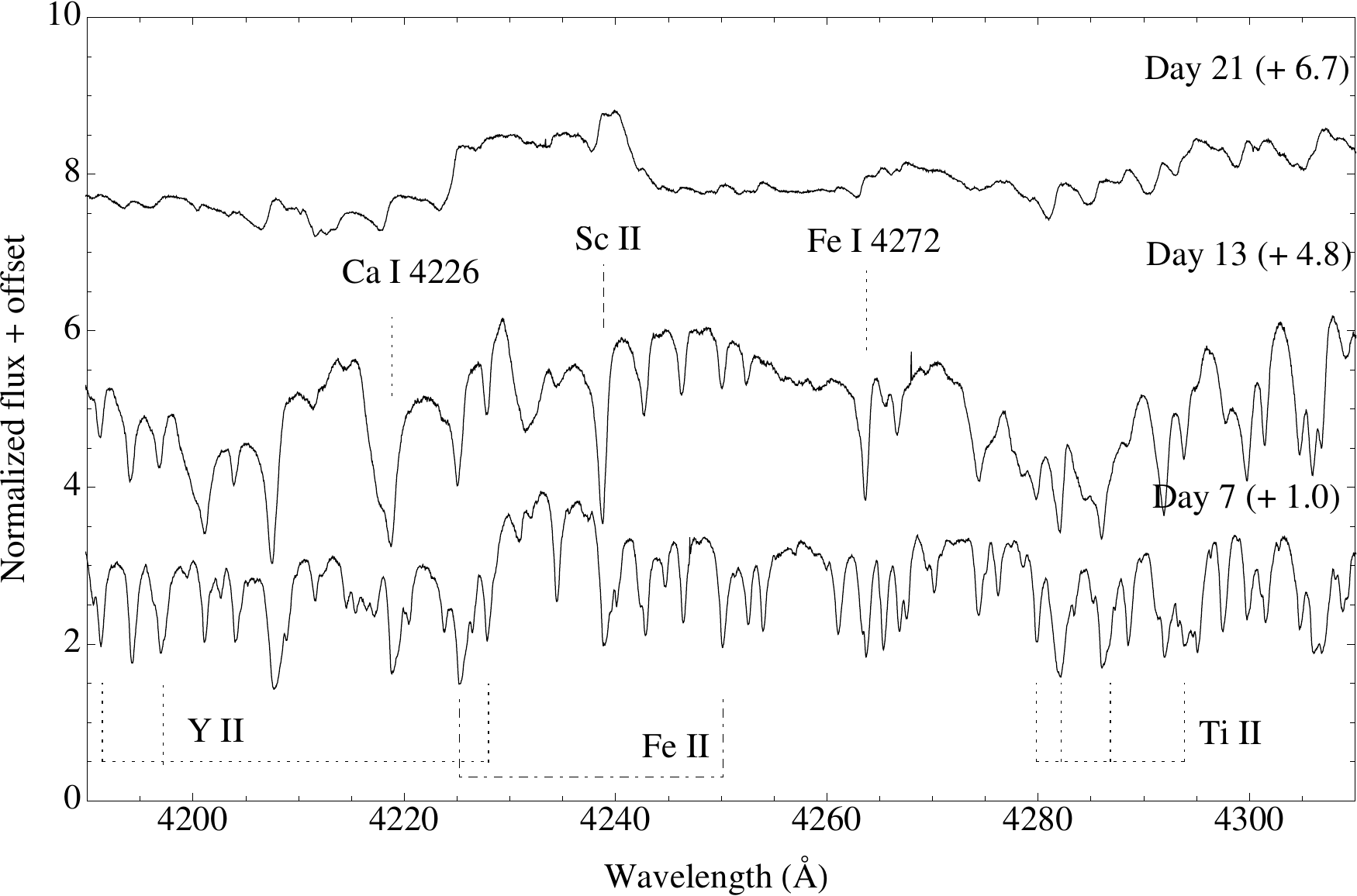}
\caption{Spectra of V1369 Cen obtained on day 7, 13 and 21 between 4190 and 4310 \AA\, with the identifications of some narrow absorptions lines.}
\label{fig:no2b}
\end{figure}

We have identified, via cross--correlation procedure between observed narrow features and laboratory wavelengths\footnote{We referred to the Atomic Line List (v 2.05) maintained by Peter van Hoof, http://www.pa.uky.edu/$\sim$peter/newpage/}, $319$ (over a total of $424$ absorptions at day 7) low-excitation ($E_{in} \leq 6$ eV) transitions of singly-ionized heavy elements (Ba, Cr, Fe, Mn, Sc, Sr, Ti, V, and Y - see for example \citealp{Williams2008}), all of them characterized by an expanding velocity of $v_{exp} \sim $ - 550 km/s, see Fig.\ref{fig:no2b}.  This procedure resulted in some non-identifications and degenerate-identifications, which we estimate to represent a small fraction (less than 20$\%$) of the entire dataset. A complete analysis will be published elsewhere (Izzo et al. in preparation).

Although from day 11 an additional expanding component is detected in the P-Cygni profiles of the Na I doublet, see right panel in Fig. \ref{fig:no1}, no further expanding components are identified for these heavy elements absorbing systems: the cross--correlation method provides a minor number of identifications considering a unique expanding velocity of $\sim$ 1100 km/s. If we consider the presence of both expanding components for these heavy elements, the number of acceptable identifications for the higher velocity components is a small fraction $\sim 1\%$ of the identified lower velocity systems. 

\begin{figure*}[!t]
\center
\includegraphics[width=0.325\textwidth]{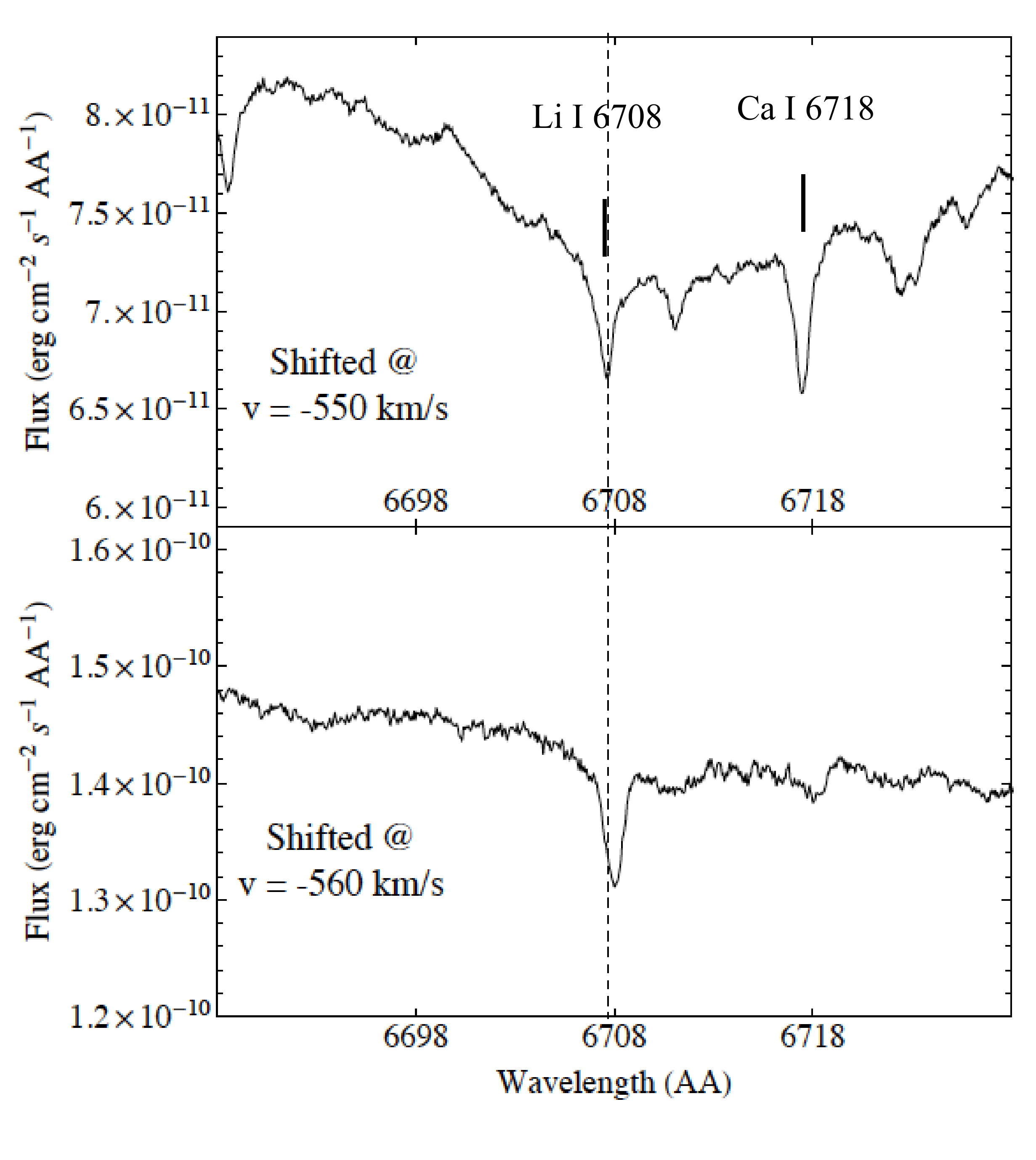}
\includegraphics[width=0.33\textwidth]{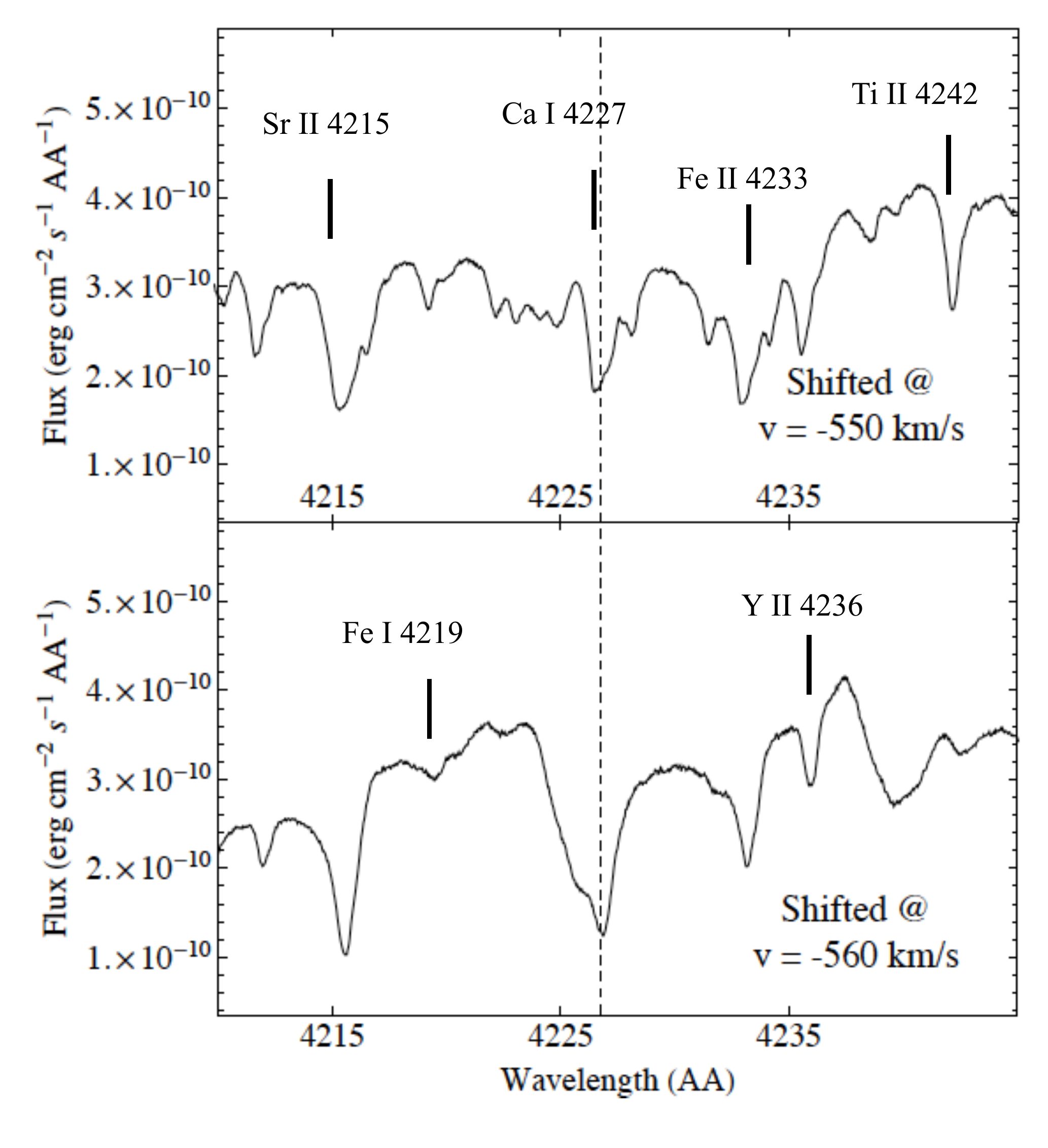}
\includegraphics[width=0.325\textwidth]{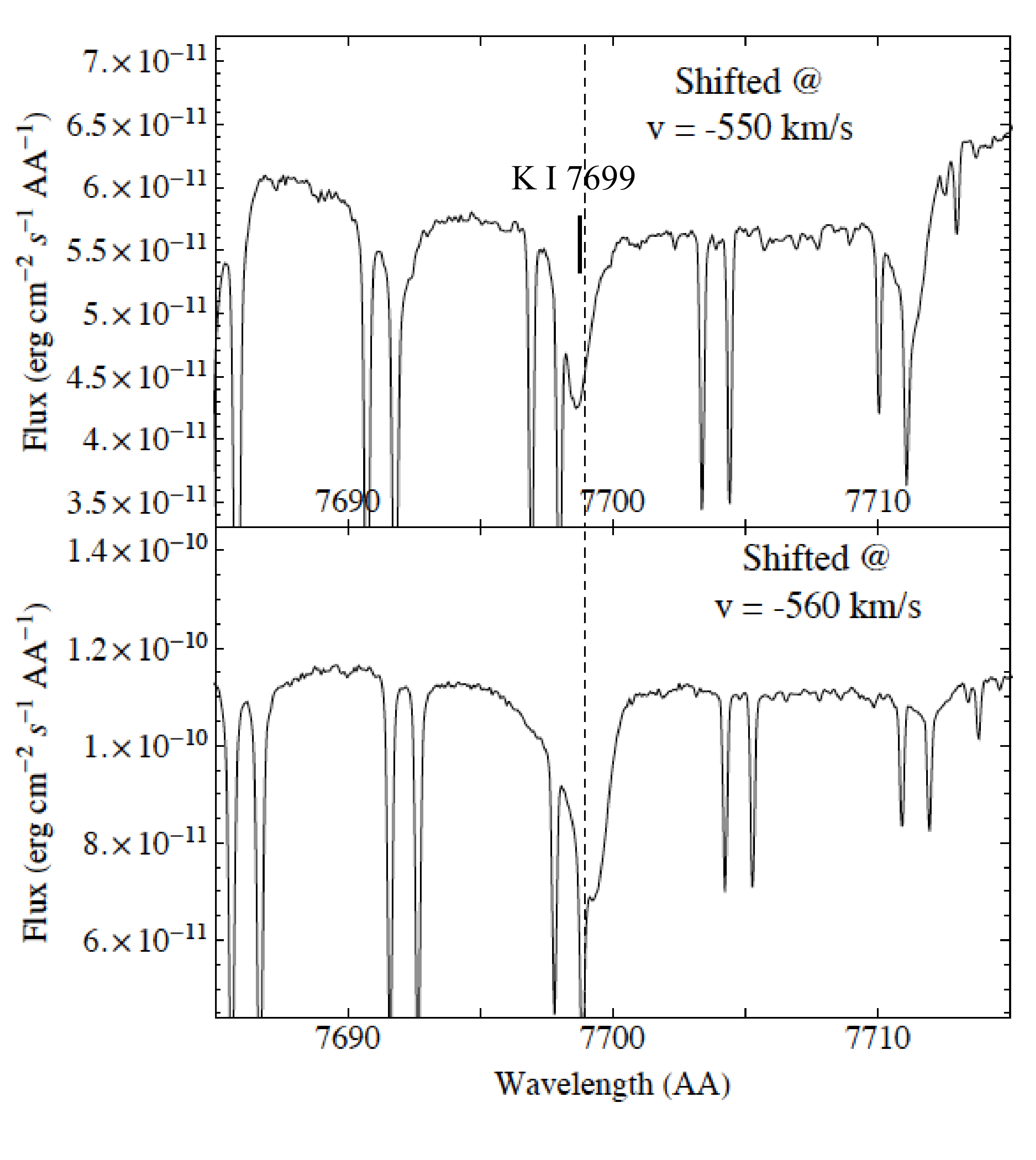}\\
\caption{The identification of Li I 6708 (left), Ca I 4227 (medium) and K I 7699 (right) features by direct comparison with the Na I D2 5890 \AA\, P--Cygni absorption in the day 7 (upper panels) and day 13 (lower panels) spectra. All the features share same expansion velocity ($v_{exp,F1} = -550$ km/s, $v_{exp,F2} = -560$ km/s) as that of Sodium. }
\label{fig:no2}
\end{figure*}

Among the many narrow transitions, we have identified many low-ionization neutral elements transitions (Fe, Ca, K), all belonging to the one expanding absorption system at the same velocity that observed in ionized heavy elements, with the only exception being Na I D lines, which show additional expanding components from day 11, see Fig. \ref{fig:no1}.  In particular, we note the clear presence of the resonance transitions of $^7$Li I 6708, Ca I 4227, and K I 7699, see Fig. \ref{fig:no2}, all of them with an expansion velocity of $v_{exp} \sim$ -550 km/s, in the spectra of day 7 to day 18.

\subsection{Possible alternative identifications}

Here below we discuss several plausible alternatives to our $^7$Li identification. 

1. We can exclude the possibility that this feature is due to a diffuse interstellar band (DIB -- \citealp{Herbig1995,Bondar2012}), because i) no known DIBs are located at 6695.6 \AA\,; ii) all observed heavy elements narrow absorptions vanish after $\sim$ 20 days from the initial outburst  when the nova continuum is still bright and DIBs should persist, being related to interstellar medium located in between the background exciting radiation and the observer; iii) we observe variations in the observed wavelengths of all these narrow absorptions, an effect that DIBs do not show.

2. An other possibility that must be considered is whether the  6695.6 \AA\, absorption line is due to metal lines excited by resonance absorption of UV radiation which is absorbed in the iron curtain phase and then reprocessed at longer wavelengths \citep{Johansson1983,ShoreBASI}. For example, each of the resonance transitions of Li, Ca, and K could be alternatively identified with a transition of Fe or an s-process element at that wavelength, assuming that they are pumped by UV radiation (either continuum or lines). Particularly for the Li I 6707.8 \AA\, resonance doublet, a possible identification could be Fe II 6707.54 \AA\,. In this case one should also expect absorption features from the same lower  energy level configuration ($3d^6(^5D)5p$), and from the same term ($4Fo-4G$), which results in additional 2 other possible transitions (Fe II 6769.27, 6811.49 \AA\,) in the observed spectra. We have checked for\footnote{we consider a $\Delta \lambda = |\lambda_{obs} - \lambda_{lab}| \leq$ 0.4 \AA.} their presence in the spectra of day 7 and day 13\footnote{these epochs corresponds to the first two FEROS observations, which show the largest number of narrow absorptions, due also to a greater resolution, in our entire spectral database},  but we did not find any absorption feature corresponding to Fe II 6769.27 and 6811.49 \AA\,. This evidence disfavours an UV-pumped origin for the absorption at 6695.6 \AA\,, even if considering the coupling between the transitions and the ejecta velocity field. Conversely, it supports our initial identification as expanding Li I 6708 \AA. 

\begin{figure}[!t]
\center
\includegraphics[width=0.49\textwidth]{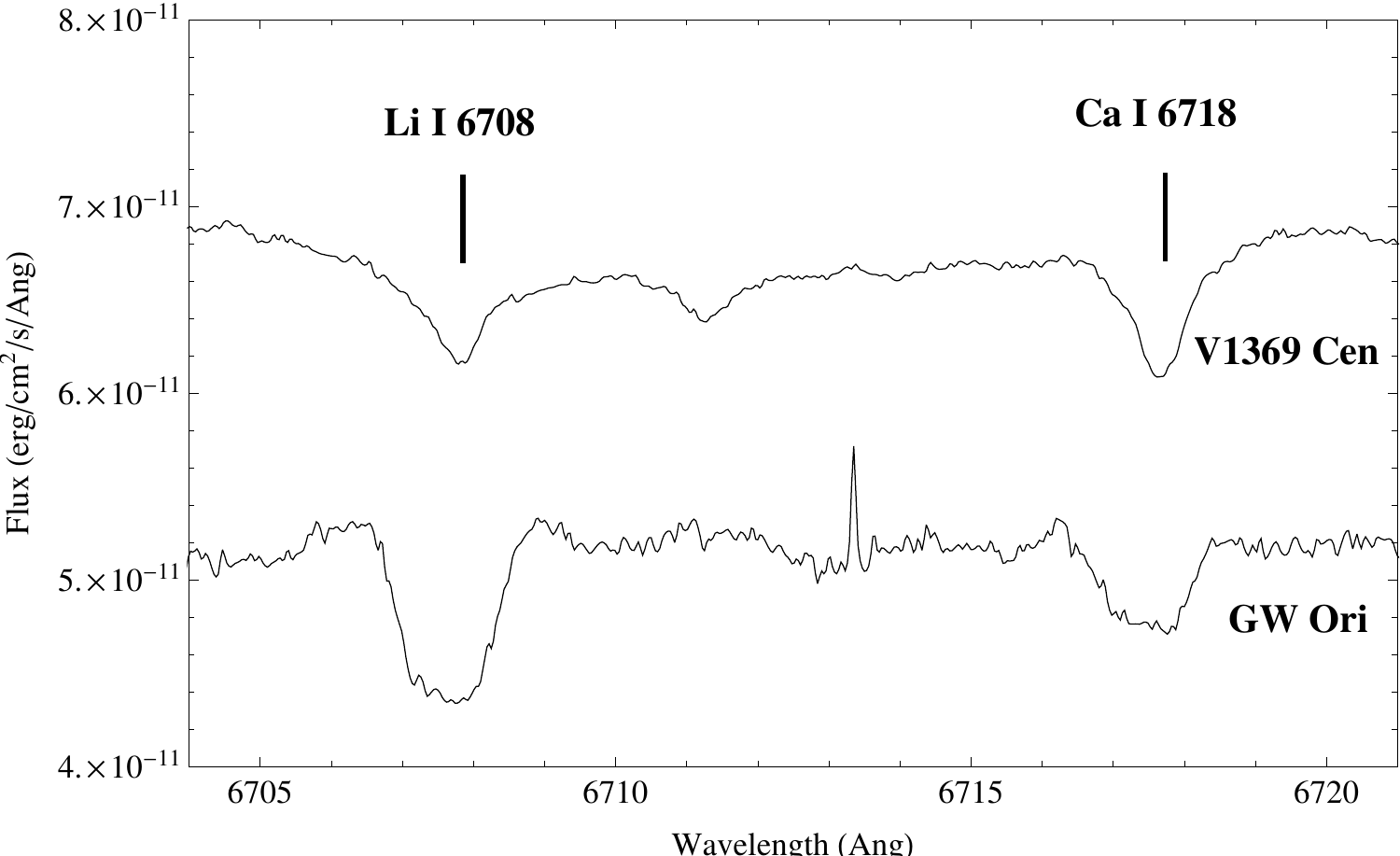}
\caption{The comparison between the spectra of the V1369 Cen and GW Ori. It is clearly evident in
both spectra the presence of $^7$Li I 6708 \AA\, absorption line, as well Ca I 6718 \AA. The absorption
in the nova spectrum around  $\lambda = 6711.3$ is identified as Cr II 6711.29 \AA.}
\label{fig:no3}
\end{figure}

3.  We compare the red spectral region of Nova Cen with the same region observed in GW Ori, a T Tauri star characterized by the presence of Lithium \citep{Bonsack1960}. GW Ori was observed with FEROS and we have selected the observation of 26 November 2010. In Fig. \ref{fig:no3} we show the comparison of these two spectra, after correcting the nova spectrum for the expanding velocity of $v_{exp} = -550$ km/s.  We clearly see the presence in the GW Ori of broad  Ca I 6718 \AA\, and Li I 6708\AA\, giving more support to the identification of absorptions at  6695.6 \AA\ and 6705.3 \AA\,  detected in V1369, with Li I 6708 and Ca I 6718.

\section{Results and Discussions}

Under the assumption that this absorption is $^7$Li, we have estimated its ejected mass following 
\citet{Friedjung1979}. Since lithium, sodium and potassium are alkali metals with very similar Grotrian diagrams and with respective resonant transitions, differing by only 0.25 eV for K-Li and Li-Na, we can assume that resonance doublets form under similar conditions. This assumption implies that the ratio of their optical thickness $\tau_i$ is related to their abundance ratio multiplied by  their respective $gf$ ratio. Following \citealp{Spitzer1998} (Eq. 3-48) we have for the case of Li/Na :

\begin{equation}\label{eq:no2}
\frac{A_m(Li)}{A_m(Na)} = \left(\frac{W_{Li6708}}{6708^2} / \frac{W_{NaD2}}{5890^2}\right) \times \frac{gf _{NaD2}}{gf _{Li6708}} \times \frac{u_{Li}}{u_{Na}},
\end{equation}
where $W_{\lambda_i}$ the measured equivalent width at the transition wavelength $\lambda_i$, and $u_{i}$ the atomic mass of the corresponding element (in our cases $u_{Li} = 7$, $u_{Na} = 23$ and $u_{K} = 39$). For the $^7$Li I doublet blend, we have considered the value of log gf = -0.174, whereas the single components has respectively log gf D1 = -0.00177, and  log gf D2 = -0.3028 \citep{Kramida2013}.
We have obtained $A_m(Li)/A_m(Na) = 3.2 / 100$ and $A_m(Li)/A_m(K) = 6.5 / 100$ on day 7, and $A_m(Li)/A_m(Na) = 2.4 / 100$ and $A_m(Li)/A_m(K) = 5.2 / 100$ on day 13. The lithium log overabundances are  4.8 and 3.9 with respect to sodium and potassium solar abundances \citep{Lodders2009}. This result implies an overabundance of lithium in the nova ejecta of the order of $10^4$, which is largely enough for explaining the Galactic $^7$Li enrichment \citep{Friedjung1979}. The mass of sodium and potassium ejected in novae can be computed in terms of solar mass by using the results of different nova composition models, characterized by different WD masses, accretion rates and mixing degree \citep{JoseHernanz1998}. We have considered the results obtained for CO nova models, where the mass of ejected sodium ranges from 3.4 $\times$ 10$^{-5}$ to 2.0 $\times$ 10$^{-4}$ the mass of ejected hydrogen, while the ejected potassium varies from 5.1 and 7.2 $\times$ 10$^{-6}$. The hydrogen ejected mass can be estimated both from the intensity of H$\beta$ when the ejecta is completely optically thin, e.g. in late nebular phases \citep{Mason2005}, and also through the observed $t_2$ value \citep{DellaValle2002}. Both methods converge toward the value of an ejected hydrogen mass of $M_{H,ej} \approx 10^{-4}$ M$_{\odot}$. After combining the Na and K ejecta measurements with the respective lithium mass abundance ratio, we obtain that the mass of lithium ejected by V1369 Cen to be in the range $M_{Li,1} =$ 0.3 -  4.8 $\times 10^{-10} $ M$_{\odot}$. 
The ratios Li/Na and Li/K have been determined on day 7 an 13 when only a small amount of $^7$Be has decayed to $^7$Li. Hence, the measured amount of $^7$Li may be only a lower limit. However, TNRs which produce $^7$Be could start even years or months (see Starrfield, Iliadis and Hix in \citealp{BodeEvans}) before the nova was discovered. Therefore the most plausible scenario for V1369 Cen is the one in which TNRs started weeks/months before the optical detection, and therefore the measured abundance of lithium can be considered a good approximation of the total amount of lithium actually produced by the nova (or a firm lower limit). 
The crucial quantity needed to compute the global Li yield of the galactic nova population is the Galactic nova rate, which is known within a factor two,  $R_N$ = 20--34 events/yr \citep{DellaValle1994,Shafter1997}. With the yield obtained above, we derive for the lithium mass injected in the Milky Way by nova systems
$M_{Li,tot} \sim$ 2 - 45 M$_{\odot}$/Gyr. However, it is well known that the ejecta of ''slow'' novae are more massive than ejecta of ''fast'' novae, by an order of magnitude, $\sim 10^{-4}$ M$_{\odot}$ vs. $\sim 10^{-5}$ M$_{\odot}$ \citep{DellaValle2002}, therefore fast novae, which can form $\sim$ 30\% of the nova population of the Milky Way \citep{DellaValle1993}, should contribute only marginally to the global $^7Li$ yield. The above reported range of lithium mass decreases to  $M_{Li,tot} \leq 17 $ M$_{\odot}$/Gyr, for a rate of ''slow'' novae of 15-24 events/year, in good agreement with the theoretical predictions \citep{JoseHernanz1998}.

\begin{figure}[!t]
\center
\includegraphics[width=0.49\textwidth]{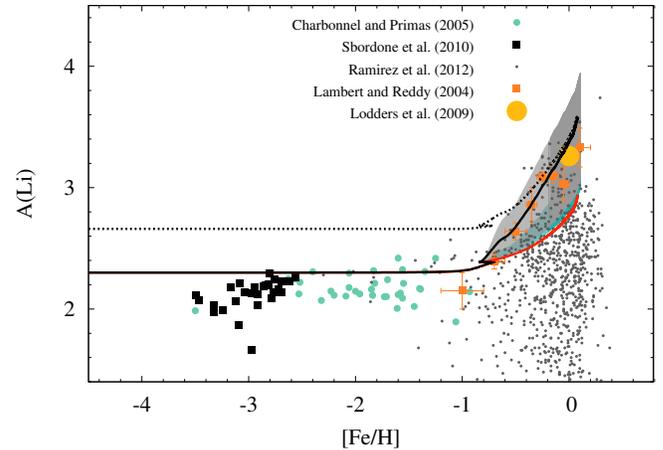}
\caption{A(Li) vs [Fe/H] for solar neighbourhood stars and meteorites (symbols -- see legend) compared to the predictions of chemical evolution models (lines and coloured areas).  The back and forth behavior in the theoretical curves around [Fe/H]=$-$0.8 is due to the transition between the halo/thick-disc and thin-disc formation phases (see text).}
\label{fig:no4}
\end{figure}

In Fig. \ref{fig:no4} it is shown the A(Li) vs. [Fe/H] observed relation compared with Galactic chemical evolution model results. The Galactic chemical evolution model used is an updated version of the chemical evolution model of \citet{Romano1999,Romano2001}, that is based on the two-infall model of \citet{Chiappini1997}, and that includes AGB stars \citep{Karakas2010}, super-AGB stars \citep{Dohertya}, Galactic cosmic rays \citep{Lemoine1998} and novae, as Li producers. In this model, the Galactic inner halo and thick disc form by accretion of gas of primordial chemical composition on a short timescale ($\sim 1$ Gyr). The gas is efficiently turned into stars as long as its density is above a critical threshold, below which the star formation stops. The thin disc forms out of a second episode of infall of gas of mainly extragalactic origin on longer timescales (7-8 Gyr in the solar neighborhood) and with lower star formation efficiency. It is worth emphasizing that the adopted mass assembly history is consistent with what is obtained for Milky Way-like galaxies in a full cosmological framework \citep{Colavitti2008}. The black continuous line in Fig. \ref{fig:no4} is the best fit to the data obtained with a model with all Li sources, starting from a primordial Li abundance of $A(Li) = 2.3$ and by assuming that each "slow" nova (the adopted current slow novae rate is 17 events/yr) ejects on an average $M_{Li} =$  2.55 $\times 10^{-10}$ M$_{\odot}$ in agreement with the measurement of $M_{Li} =$ 0.3 -  4.8 $\times 10^{-10} $ M$_{\odot}$ presented in this paper. The black dashed line shows the predictions of the same model when a high primordial Li abundance is adopted (see the Introduction). The red line is the best model with all Li factories but novae: it is clearly seen that novae are necessary to explain the late rise from the plateau value. The grey area indicates the uncertainties in the model predictions due to uncertainties in both the estimated Li yield from novae and the current slow nova rate: the upper (lower) boundary refers to the upper (lower) limit to the Li yield estimated in this paper and a maximum (minimum) current slow nova rate of 24 (15) events/yr. The light green area similarly indicates the uncertainties in the model predictions when the maximum and minimum Li yields from \cite{JoseHernanz1998} are assumed. Though the two areas partly overlap, it is clearly seen that the theoretical nova yields tend to underproduce Li in the Galaxy, while the semi-empirical yields estimated in this paper give a better match with observed data points. Should the result presented here be confirmed by further observations of $^7$Li, Classical novae would stand as one of the major Li producers on a Galactic scale.

\acknowledgments

We thank the referee for her/his constructive comments/criticisms which have improved the paper and Steven Shore, Marina Orio and Paolo Molaro for useful discussions. We are grateful to Roland Gredel for DDT programme 091.A-9032 B. We acknowledge support by project Fondecyt n. 1130849.





\begin{thebibliography}{}
\expandafter\ifx\csname natexlab\endcsname\relax\def\natexlab#1{#1}\fi

\bibitem[{{Bode} \& {Evans}(2008)}]{BodeEvans}
{Bode}, M.~F., \& {Evans}, A. 2008, {Classical Novae.}

\bibitem[{{Bondar}(2012)}]{Bondar2012}
{Bondar}, A. 2012, MNRAS, 423, 725

\bibitem[{{Bonifacio} {et~al.}(2007){Bonifacio}, {Molaro}, {Sivarani},
  {Cayrel}, {Spite}, {Spite}, {Plez}, {Andersen}, {Barbuy}, {Beers}, {Depagne},
  {Hill}, {Fran{\c c}ois}, {Nordstr{\"o}m}, \& {Primas}}]{Bonifacio2007}
{Bonifacio}, P., {Molaro}, P., {Sivarani}, T., {et~al.} 2007, A$\&$A, 462, 851

\bibitem[{{Bonsack} \& {Greenstein}(1960)}]{Bonsack1960}
{Bonsack}, W.~K., \& {Greenstein}, J.~L. 1960, ApJ, 131, 83

\bibitem[{{Brandi} {et~al.}(2009){Brandi}, {Quiroga}, {Miko{\l}ajewska},
  {Ferrer}, \& {Garc{\'{\i}}a}}]{Brandi2009}
{Brandi}, E., {Quiroga}, C., {Miko{\l}ajewska}, J., {Ferrer}, O.~E., \&
  {Garc{\'{\i}}a}, L.~G. 2009, A$\&$A, 497, 815

\bibitem[{{Cameron} \& {Fowler}(1971)}]{CameronFowler1971}
{Cameron}, A.~G.~W., \& {Fowler}, W.~A. 1971, ApJ, 164, 111

\bibitem[{{Charbonnel} \& {Primas}(2005)}]{Charbonnel2005}
{Charbonnel}, C., \& {Primas}, F. 2005, A$\&$A, 442, 961

\bibitem[{{Chiappini} {et~al.}(1997){Chiappini}, {Matteucci}, \&
  {Gratton}}]{Chiappini1997}
{Chiappini}, C., {Matteucci}, F., \& {Gratton}, R. 1997, ApJ, 477, 765

\bibitem[{{Coc} {et~al.}(2014){Coc}, {Uzan}, \& {Vangioni}}]{Coc2014}
{Coc}, A., {Uzan}, J.-P., \& {Vangioni}, E. 2014, JCAP, 10, 50

\bibitem[{{Colavitti} {et~al.}(2008){Colavitti}, {Matteucci}, \&
  {Murante}}]{Colavitti2008}
{Colavitti}, E., {Matteucci}, F., \& {Murante}, G. 2008, A$\&$A, 483, 401

\bibitem[{{D'Antona} \& {Matteucci}(1991)}]{DAntonaMatteucci1991}
{D'Antona}, F., \& {Matteucci}, F. 1991, A$\&$A, 248, 62

\bibitem[{{Della Valle} \& {Duerbeck}(1993)}]{DellaValle1993}
{Della Valle}, M., \& {Duerbeck}, H.W. 1993, A$\&$A, 271, 175

\bibitem[{{Della Valle} \& {Livio}(1994)}]{DellaValle1994}
{Della Valle}, M., \& {Livio}, M. 1994, A$\&$A, 286, 786

\bibitem[{{Della Valle} \& {Livio}(1995)}]{DellaValle1995}
{Della Valle}, M., \& {Livio}, M. 1995, ApJ, 452, 704

\bibitem[{{Della Valle} {et~al.}(2002){Della Valle}, {Pasquini}, {Daou}, \&
  {Williams}}]{DellaValle2002}
{Della Valle}, M., {Pasquini}, L., {Daou}, D., \& {Williams}, R.~E. 2002,
  A$\&$A, 390, 155

  \bibitem[{{Doherty} {et~al.}(2014a){Doherty}, {Gil-Pons}, {Lau}, {Lattanzio}, \& {Siess}}]{Dohertya}
{Doherty}, C.~L., {Gil-Pons}, P., {Lau}, H.~H.~B., {et~al.} 2014a, MNRAS, 441,
  195

\bibitem[{{Friedjung}(1979)}]{Friedjung1979}
{Friedjung}, M. 1979, A$\&$A, 77, 357

\bibitem[{{Herbig}(1995)}]{Herbig1995}
{Herbig}, G.~H. 1995, ARA$\&$A, 33, 19

\bibitem[{{Iben}(1973)}]{Iben1973}
{Iben}, Jr., I. 1973, ApJ, 185, 209

\bibitem[{{Iocco} {et~al.}(2009){Iocco}, {Mangano}, {Miele}, {Pisanti}, \&
  {Serpico}}]{Iocco2009}
{Iocco}, F., {Mangano}, G., {Miele}, G., {Pisanti}, O., \& {Serpico}, P.~D.
  2009, Phys. Rep., 472, 1

\bibitem[{{Johansson}(1983)}]{Johansson1983}
{Johansson}, S. 1983, \mnras, 205, 71P

\bibitem[{{Jos{\'e}} \& {Hernanz}(1998)}]{JoseHernanz1998}
{Jos{\'e}}, J., \& {Hernanz}, M. 1998, ApJ, 494, 680

\bibitem[{{Karakas}(2010)}]{Karakas2010}
{Karakas}, A.~I. 2010, MNRAS, 403, 1413

\bibitem[{{Kaufer} {et~al.}(1999){Kaufer}, {Stahl}, {Tubbesing},
  {N{\o}rregaard}, {Avila}, {Francois}, {Pasquini}, \& {Pizzella}}]{Kaufer1999}
{Kaufer}, A., {Stahl}, O., {Tubbesing}, S., {et~al.} 1999, The Messenger, 95, 8

\bibitem[{{Kolb} \& {Turner}(1990)}]{KolbTurner}
{Kolb}, E.~W., \& {Turner}, M.~S. 1990, {The early universe.}

\bibitem[{{Korn} {et~al.}(2006){Korn}, {Grundahl}, {Richard}, {Barklem},
  {Mashonkina}, {Collet}, {Piskunov}, \& {Gustafsson}}]{Korn2006}
{Korn}, A.~J., {Grundahl}, F., {Richard}, O., {et~al.} 2006, Nature, 442, 657

\bibitem[{{Kramida} {et~al.}(2013){Kramida}, {Ralchenko}, {Reader} et al.}]{Kramida2013}
{Kramida}, A., {Ralchenko}, Y., {Reader}, L.,  \& the NIST ASD team, 2013, NIST Atomic 
Spectra Database Ver. 5.1 http://physics.nist.gov/asd

\bibitem[{{Lambert} \& {Reddy}(2004)}]{Lambert2004}
{Lambert}, D. L., \& {Reddy}, B. E., 2004, \mnras, 349, 757L

\bibitem[{{Lemoine} {et~al.}(1998){Lemoine}, {Vangioni-Flam}, \&
  {Cass{\'e}}}]{Lemoine1998}
{Lemoine}, M., {Vangioni-Flam}, E., \& {Cass{\'e}}, M. 1998, ApJ, 499, 735

\bibitem[{{Livio} \& {Truran}(1987)}]{Livio1987}
{Livio}, M., \& {Truran}, J. W., 1987, ApJ, 318, 316

\bibitem[{{Lodders} {et~al.}(2009){Lodders}, {Palme}, \& {Gail}}]{Lodders2009}
{Lodders}, K., {Palme}, H., \& {Gail}, H.-P. 2009, Landolt B{\"o}rnstein, 44

\bibitem[{{Mason} {et~al.}(2005){Mason}, {Della Valle}, {Gilmozzi}, {Lo Curto},
  \& {Williams}}]{Mason2005}
{Mason}, E., {Della Valle}, M., {Gilmozzi}, R., {Lo Curto}, G., \& {Williams},
  R.~E. 2005, A$\&$A, 435, 1031

\bibitem[{{Prantzos}(2012)}]{Prantzos2012}
{Prantzos}, N. 2012, A$\&$A, 542, A67

\bibitem[{{Prialnik} \& {Kovetz}(1995)}]{Prialnik1995}
{Prialnik}, D., \& {Kovetz}, A. 1995, ApJ, 445, 789

\bibitem[{{Romano} {et~al.}(1999){Romano}, {Matteucci}, {Molaro}, \&
  {Bonifacio}}]{Romano1999}
{Romano}, D., {Matteucci}, F., {Molaro}, P., \& {Bonifacio}, P. 1999, A$\&$A,
  352, 117

\bibitem[{{Romano} {et~al.}(2001){Romano}, {Matteucci}, {Ventura}, \&
  {D'Antona}}]{Romano2001}
{Romano}, D., {Matteucci}, F., {Ventura}, P., \& {D'Antona}, F. 2001, A$\&$A,
  374, 646
  
  \bibitem[{{Sbordone} {et~al.}(2010){Sbordone}, {Bonifacio}, {Caffau}, {Ludwig}, 
  {Behara}, {Gonz{\'a}lez Hern{\'a}ndez}, 
  {Steffen}, {Cayrel}, {Freytag}, {van't Veer}, {Molaro}, {Plez}, {Sivarani}, {Spite}, 
{Spite}, {Beers}, {Christlieb}, {Fran{\c c}ois}, {Hill}}]{Sbordone2010}
{Sbordone}, L., {Bonifacio}, P., {Caffau}, E., {et~al.} 2010, A$\&$A, 522, 26

\bibitem[{{Shafter}(1997)}]{Shafter1997}
{Shafter}, A.~W. 1997, ApJ, 487, 226

\bibitem[{{Shore}(2012)}]{ShoreBASI}
{Shore}, S.~N. 2012, Bulletin of the Astronomical Society of India, 40, 185

\bibitem[{{Shore} {et~al.}(2011){Shore}, {Wahlgren}, {Augusteijn}, {Liimets},
  {Page}, {Osborne}, {Beardmore}, {Koubsky}, {{\v S}lechta}, \&
  {Votruba}}]{Shore2011b}
{Shore}, S.~N., {Wahlgren}, G.~M., {Augusteijn}, T., {et~al.} 2011, A$\&$A,
  527, A98

\bibitem[{{Spite} \& {Spite}(1982)}]{Spite1982}
{Spite}, F., \& {Spite}, M. 1982, A$\&$A, 115, 357

\bibitem[{{Spitzer}(1998)}]{Spitzer1998}
{Spitzer}, L.~J., 1998, {Physical Processes in the Interstellar Medium.}

\bibitem[{{Starrfield} {et~al.}(1978){Starrfield}, {Truran}, {Sparks}, \&
  {Arnould}}]{Starrfield1978}
{Starrfield}, S., {Truran}, J.~W., {Sparks}, W.~M., \& {Arnould}, M. 1978, ApJ,
  222, 600

\bibitem[{{Tajitsu} {et~al.}(2015){Tajitsu}, {Sadakane}, {Naito}, {Arai}, \&
  {Aoki}}]{Tajitsu2015}
{Tajitsu}, A., {Sadakane}, K., {Naito}, H., {Arai}, A., \& {Aoki}, W. 2015,
  Nature, 518, 381

\bibitem[{{Travaglio} {et~al.}(2001){Travaglio}, {Randich}, {Galli},
  {Lattanzio}, {Elliott}, {Forestini}, \& {Ferrini}}]{Travaglio2001}
{Travaglio}, C., {Randich}, S., {Galli}, D., {et~al.} 2001, ApJ, 559, 909

\bibitem[{{Truran}(1981)}]{Truran1981}
{Truran}, J.~W. 1981, Progress in Particle and Nuclear Physics, 6, 177

\bibitem[{{Vanzi} {et~al.}(2012){Vanzi}, {Chacon}, {Helminiak}, {Baffico},
  {Rivinius}, {{\v S}tefl}, {Baade}, {Avila}, \& {Guirao}}]{Vanzi2012}
{Vanzi}, L., {Chacon}, J., {Helminiak}, K.~G., {et~al.} 2012, MNRAS, 424, 2770

\bibitem[{{Ventura} \& {D'Antona}(2010)}]{Ventura2010}
{Ventura}, P., \& {D'Antona}, F. 2010, MNRAS, 402, L72

\bibitem[{{Wallerstein} {et~al.}(2008){Wallerstein}, {Harrison}, {Munari}, \&
  {Vanture}}]{Wallerstein2008}
{Wallerstein}, G., {Harrison}, T., {Munari}, U., \& {Vanture}, A. 2008, PASP,
  120, 492

\bibitem[{{Williams} {et~al.}(2008){Williams}, {Mason}, {Della Valle}, \&
  {Ederoclite}}]{Williams2008}
{Williams}, R., {Mason}, E., {Della Valle}, M., \& {Ederoclite}, A. 2008, ApJ,
  685, 451

\end{thebibliography}
\end{document}